\documentclass{article}
\usepackage{arxiv}
\usepackage[utf8]{inputenc} 
\usepackage[T1]{fontenc}    
\usepackage{hyperref}       
\usepackage{url}         
\usepackage{booktabs}    
\usepackage{amsfonts}      
\usepackage{nicefrac}      
\usepackage{microtype}     
\usepackage{lipsum}
\usepackage{float}
\usepackage{graphicx}
\usepackage{caption}
\usepackage{subcaption}
\usepackage{titlesec}
\usepackage{multirow}
\usepackage{amsmath}
\usepackage{soul,color}
\usepackage[section]{placeins}

\usepackage{array, makecell} %
\usepackage{float}
\usepackage[section]{placeins}

\usepackage[nodisplayskipstretch]{setspace}

\title{Distance Matrix based Crystal Structure Prediction using Evolutionary Algorithms}
\author{
Jianjun Hu*\\
  Department of Computer Science and Engineering\\
  University of South Carolina\\
  Columbia, SC 29201 \\
\texttt{jianjunh@cse.sc.edu} 
  \And
 Wenhui Yang\\
 School of Mechanical Engineering\\
  Guizhou University \\
  Guiyang China 550050 \\
  \And
    Edirisuriya M. Dilanga Siriwardane\\
  Department of Computer Science and Engineering\\
  University of South Carolina\\
  Columbia, SC 29201 \\
}
\titlespacing {\section}
{0pt}{5.5ex plus 1ex minus .2ex}{3.3ex plus .2ex}
\begin{document}
\maketitle
\begin{abstract}
Crystal structure prediction (CSP) for inorganic materials is one of the central and most challenging problems in materials science and computational chemistry. This problem can be formulated as a global optimization problem in which global search algorithms such as genetic algorithms (GA) and particle swarm optimization have been combined with first principle free energy calculations to predict crystal structures given only a material composition or only a chemical system. These DFT based ab initio CSP algorithms are computationally demanding and can only be used to predict crystal structures of relatively small systems. The vast coordinate space plus the expensive DFT free energy calculations limits their inefficiency and scalability. On the other hand, a similar structure prediction problem has been intensively investigated in parallel in the protein structure prediction community of bioinformatics, in which the dominating predictors are knowledge based approaches including homology modeling and threading that exploit known protein structures. Surprisingly, the CSP field has mainly focused on ab initio approaches in the past decade. Inspired by the knowledge-rich protein structure prediction approaches, herein we explore whether known geometric constraints such as the pairwise atomic distances of a target crystal material can help predict/reconstruct its structure given its space group and lattice information. We propose DMCrystal, a genetic algorithm based crystal structure reconstruction algorithm based on predicted atomic pairwise distances. Based on extensive experiments, we show that the predicted distance matrix can dramatically help to reconstruct the crystal structure and usually achieves much better performance than CMCrystal, an atomic contact map based crystal structure prediction algorithm. This implies that knowledge of atomic interaction information learned from existing materials can be used to significantly improve the crystal structure prediction in terms of both speed and quality.

\end{abstract}
\keywords{crystal structure prediction \and distance matrix \and machine learning \and contact map \and global optimization \and genetic algorithms}

\section{Introduction}








Discovery of novel materials has big potentials in transforming a variety of industries such as cell phones, electric vehicles,energy storage, quantum computing, and chemical catalysts \cite{oganov2019structure}. Compared to traditional Edisonian experimental methods which usually strongly depends on the expertise of the scientists, computational materials discovery has the advantage of much more efficient search in the vast chemical design space. In the past few years, three main strategies have emerged as the most promising approaches for new materials discovery including inverse design \cite{zunger2018inverse,kim2020inverse}, generative machine learning models \cite{dan2019generative,bradshaw2019model,kim2020inverse,noh2019inverse,ren2020inverse}, and crystal structure predictions \cite{glass2006uspex,oganov2019structure,kvashnin2019computational}.


In a typical crystal structure prediction (CSP) problem \cite{lyakhov2013new}, the goal is to find a stable and ideally synthesizable crystal structure with the lowest free energy for a given chemical composition (or a chemical system such as Mg-Mn-O with variable compositions) at given pressure–temperature conditions
\cite{oganov2019structure}. This is usually done by trying to adjust the coordinates of the given set of atoms while minimizing the free energy of the system. These kind of energy guided structure prediction approaches have been called as ab initio structure prediction. It is assumed that lower free energy corresponds to the more stable arrangement of atoms. 

With the crystal structure of a chemical compound, many physichochemical properties can be predicted reliably and routinely using first-principle calculation or machine learning models \cite{louis2020global}. The CSP approach for new materials discovery is especially appealing due to the efficient sampling algorithm that generates diverse chemically valid candidate materials compositions/formulas with low free energies \cite{dan2019generative}.  In the past 10 years, crystal structure prediction algorithms based on evolutionary algorithms and particle swarm optimization have led to a series of new materials discoveries  \cite{oganov2011evolutionary,oganov2019structure,wang2020calypso}. However, these ab initio free energy based global search algorithms have a major challenge that limits their successes to relative simple crystals  \cite{oganov2019structure,zhang2017materials} (mostly binary materials with less than 20 atoms in the unit cell \cite{oganov2019structure,wang2020calypso}) due to their dependence on the costly DFT calculations of free energies for sampled structures. With limited DFT calculations budget, it is a key issue to efficiently sample the atom configurations.  \cite{oganov2011evolutionary,lyakhov2013new}. To improve the sampling efficiency, a variety of strategies have been proposed such as exploiting symmetry \cite{chen2019tgmin,pretti2020symmetry} and pseudosymmetry \cite{lyakhov2013new}, smart variation operators, clustering\cite{sorensen2018accelerating}, machine-learning interatomic potentials with active learning  \cite{podryabinkin2019accelerating}, designing chemically based swapping operators  \cite{sharp2020chemically}. However, the scalability of these ab initio approaches remains an unsolved issue.


Recently, generative machine learning models have been emerging as a novel approach to generate new materials including generative adversarial networks (GAN) approach for both chemical composition discovery \cite{dan2020generative} and crystal structure generation for a given chemical system \cite{kim2020generative}. Compared to global free energy optimization based approaches in CSP, these methods can take advantage of the implicit compositional and atomic configuration rules and constraints embodied in the large number of known crystal structures which can be learned by the deep neural network models. Using neural networks to implicitly learn such rules can lead to more efficient sampling of the search space \cite{dan2020generative}. Actually, the idea of exploiting knowledge in existing structures is not new and has been widely used in the field protein structure prediction(PSP)  \cite{wei2019protein,zhu2017efficient,kuhlman2019advances,ryan2018crystal} and also in crystal structure prediction\cite{ryan2018crystal}. PSP algorithms can be largely divided into three categories: ab initio methods, homology modeling, and threading. Ab initio methods include energy or fragment based methods and also evolutionary covariation based approaches for contact map prediction and subsequent protein structure reconstruction. The latter two strategies are called comparative modeling \cite{nikolaev2018comparative}, in which the homology modelling is based on the fact that homologous proteins will share similar structures. The protein threading approach uses a scoring function to compare a query protein sequence to the protein structure database to identify a protein structure template. While crystal structures have no coevolutionary history, the very shared chemical assembling rules and atomic interaction rules make that many crystals of the some class share a very similar structures such as ABO\textsubscript{3} perovskites and many oxides of the same space groups. 

Herein we propose a new knowledge-rich approach for distance matrix based crystal structure prediction, which is inspired by the recent success of deep learning approaches for protein structure prediction (PSP) \cite{zheng2019deep} in AlphaFold \cite{senior2020improved} which reconstructs the protein structure from a predicted distance matrix. In the protein structure prediction problem, one has to predict the 3D tertiary structure of a protein given only its amino acid sequence. The latest approach uses deep learning to predict the contact maps \cite{emerson2017protein} or distance matrix \cite{senior2020improved}, which can then be used to reconstruct the full three-dimensional (3D) protein structure with high accuracy \cite{vendruscolo1997recovery}. In our prior work \cite{hu2020contact}, we have shown that given a crystal's space group, its lattice constants, and the contact map, all of which can be predicted, we can reconstruct the crystal structure atom coordinates using global optimization algorithms such as genetic algorithms, particle swarm optimization or Bayesian optimization. However, it is known that pairwise distances between atoms within a crystal structure usually are conserved across different compounds so that machine learning models can be built to predict the distance matrix of a given composition. In this paper, we are exploring how we can use global optimization algorithms to reconstruct the atomic configuration for a given composition based on its space group and the distance matrix. The idea is that we can exploit the rich atom interaction distribution or other geometric patterns or motifs  \cite{zhu2017efficient} existing in the large number of known crystal structures to predict the atomic distance matrix. Compared to atomic contact map, the distance has much more information which should allow better crystal structure reconstruction. While in this paper, we use the true space group, the lattice constants, and the distance matrix derived from known materials structures, the space group of crystal structures can be predicted using a variety of prediction algorithms  \cite{zhao2020machine,liang2020cryspnet} or be inferred from domain knowledge  \cite{cabeza2007space}. In \cite{liang2020cryspnet}, the top-3 accuracy for space group prediction ranges from 81\% to 100\% given its Bravais lattice, which can also be predicted using composition features with up to 84\% accuracy. With the given distance matrix and the space group, we investigate whether global optimization algorithms such as GAs and CMA-ES methods can be used for reconstructing/predicting its crystal structure.

Our contributions can be summarized as follows:

\begin{itemize}
  \item We propose a new knowledge rich approach for crystal structure prediction by using the pairwise atomic distance magtrix as a knowledge-rich strategy for solving the CSP problem. Our method is inspired by the success of corresponding algorithm in protein structure prediction, the AlphaFold.
  \item We define a series of benchmark test cases for testing global optimization algorithms to reconstruct the atomic configurations from atomic pairwise distances. 
  \item we conduct extensive evaluations of how global optimization algorithms perform in distance matrix based crystal structure prediction, how the errors of the predicted distances may affect the reconstruction performance. We also confirm some of the predicted structures using DFT calculation of the formation energy of the predicted structures.
\end{itemize}

\section{Materials and Methods}

\subsection{Problem formulation: distance matrix based crystal structure reconstruction}

\begin{figure}[h]
	\centering
	\begin{subfigure}{.45\textwidth}
		\includegraphics[width=\textwidth]{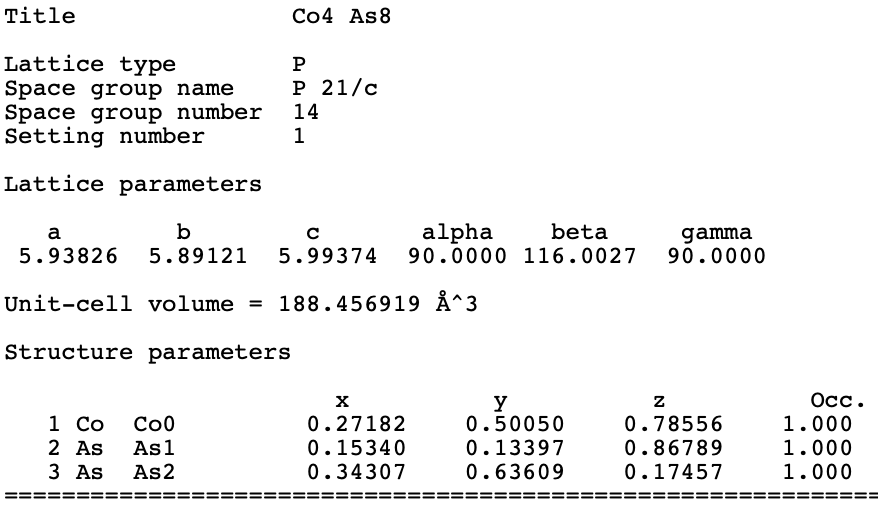}
		\caption{Cif crysal structure file}
		\vspace{3pt}
	\end{subfigure}
	\begin{subfigure}{.45\textwidth}
		\includegraphics[width=0.6\textwidth]{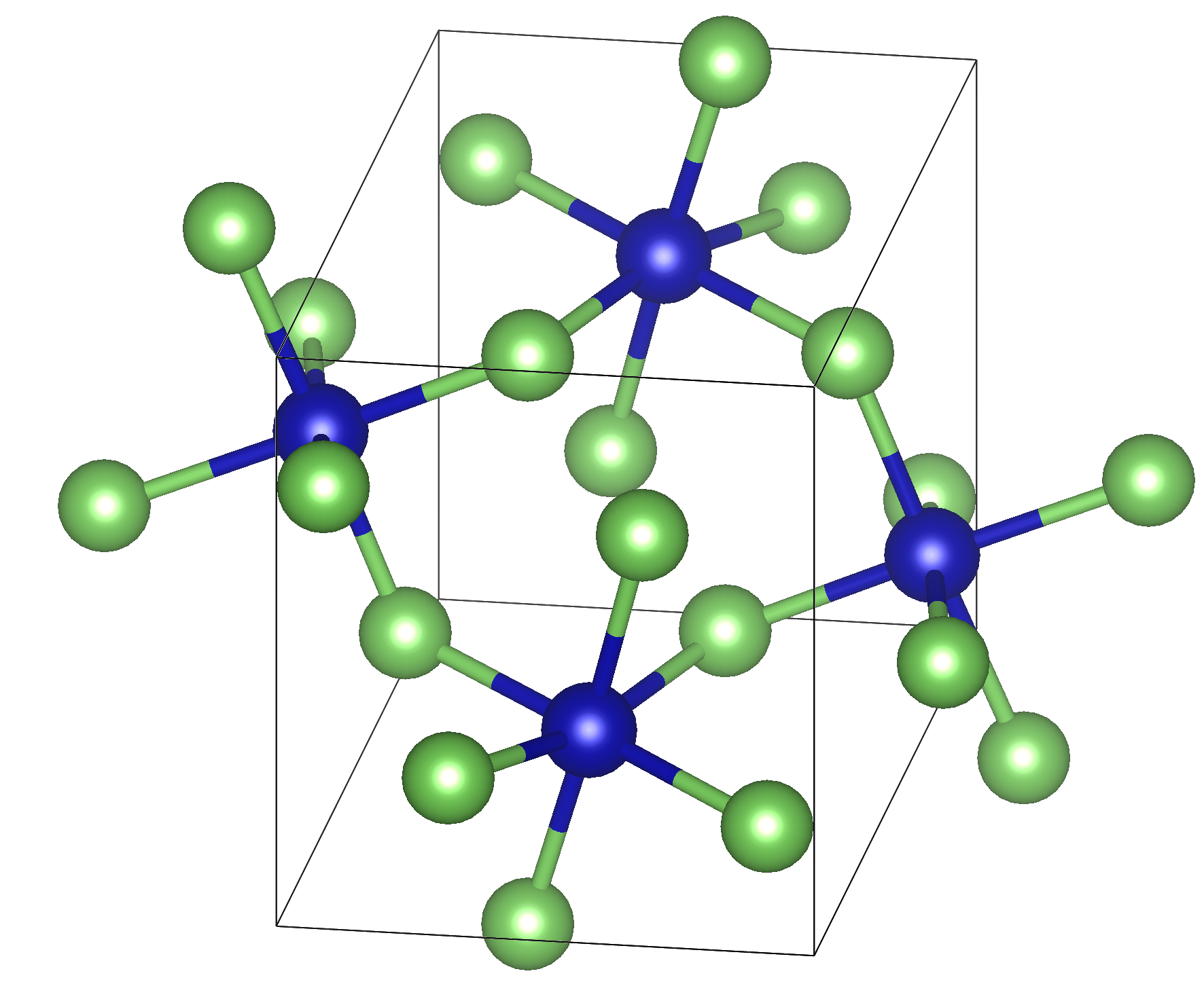}
		\caption{Graph representation}
		\vspace{3pt}
	\end{subfigure}
	\caption{Cif and graph representation of crystal material Co\textsubscript{4}As\textsubscript{8}. (a) A cif file species the space group, the six lattice parameters (a,b,c,alpha,beta,gamma, and the fractional coordinates of each site of a crystal structure.(b) a crystal structure can be represented as a graph with bond connections.}
	\label{fig:crystal_structure}
\end{figure}


A periodic crystal structure can be represented by its lattice constants a,b,c and angles $\alpha$, $\beta$, and $\gamma$, the space group, and the coordinates at unique Wyckoff positions. Using threshold for different atom pairs, the crystal structure can be converted into a graph, which can be represented as an adjacency matrix. The structure can also be looked as a point cloud, from which a pairwise distance can be derived or predicted based on known atom pair distance distribution. The pairwise distance matrix captures the interactions among atoms in the unit cell, which can be predicted by the know interaction patterns of these atom pairs in other known crystal materials structures. In this paper, we focus on the crystal structure reconstruction algorithm and assume that the pairwise atomic distance matrix has been obtained or predicted, and we'd like to check if the global optimization algorithms can help reconstruct the crystal structures in terms of the atom coordinates from the distance matrix, with or without adding other geometric or physical constraints. By formulating the distance matrix based CSP as an optimization problem, this allows us to evaluate how genetic algorithms (GA) can solve this optimization based structure reconstruction problem and how difficult this reconstruction problem for different crystal structures of varying complexity in terms of the number of unique Wyckoff positions(which determines the number of independent variables to be optimized), the level of symmetry as represented by the space group, and also the number of atoms in the unit cell. For the example in Figure~\ref{fig:crystal_structure}, the number of variables to optimize is 4x3=12, corresponding to 4 Wyckoff positions each with x,y,z three coordinate values. The crystal has 24 atoms in the unit cell, which can be mapped into a 24x24 distance matrix. The optimization problem is then how to search appropriate Wyckoff position atom coordinates so that after symmetry operations specified by space group 190, the generated crystal structure will have the same distance matrix. In this study, we assume the space group information, and the unit cell parameters of the target composition are all known, which is reasonable as they can be predicted using different approaches \cite{song2020machine,liang2020cryspnet,jiang2006prediction,nait2020prediction}. While only distance matrix information is used as optimization target, other atomic interaction information such as limits of distances or preferential neighborhood relationships (e.g. atoms of some element pairs cannot stay too close to each other in known crystals) between some atom pairs can also be added as constraints in global search. The geometric constraint optimization objective can also be combined with the traditional free energy objective to achieve synergistic effect by e.g. reducing the number of DFT free energy calculations. Our work is in principle to the reconstruction algorithm used in contact map in protein structure prediction  \cite{vendruscolo1997recovery}.

\begin{figure}[ht]
  \centering
  \includegraphics[width=0.85\linewidth]{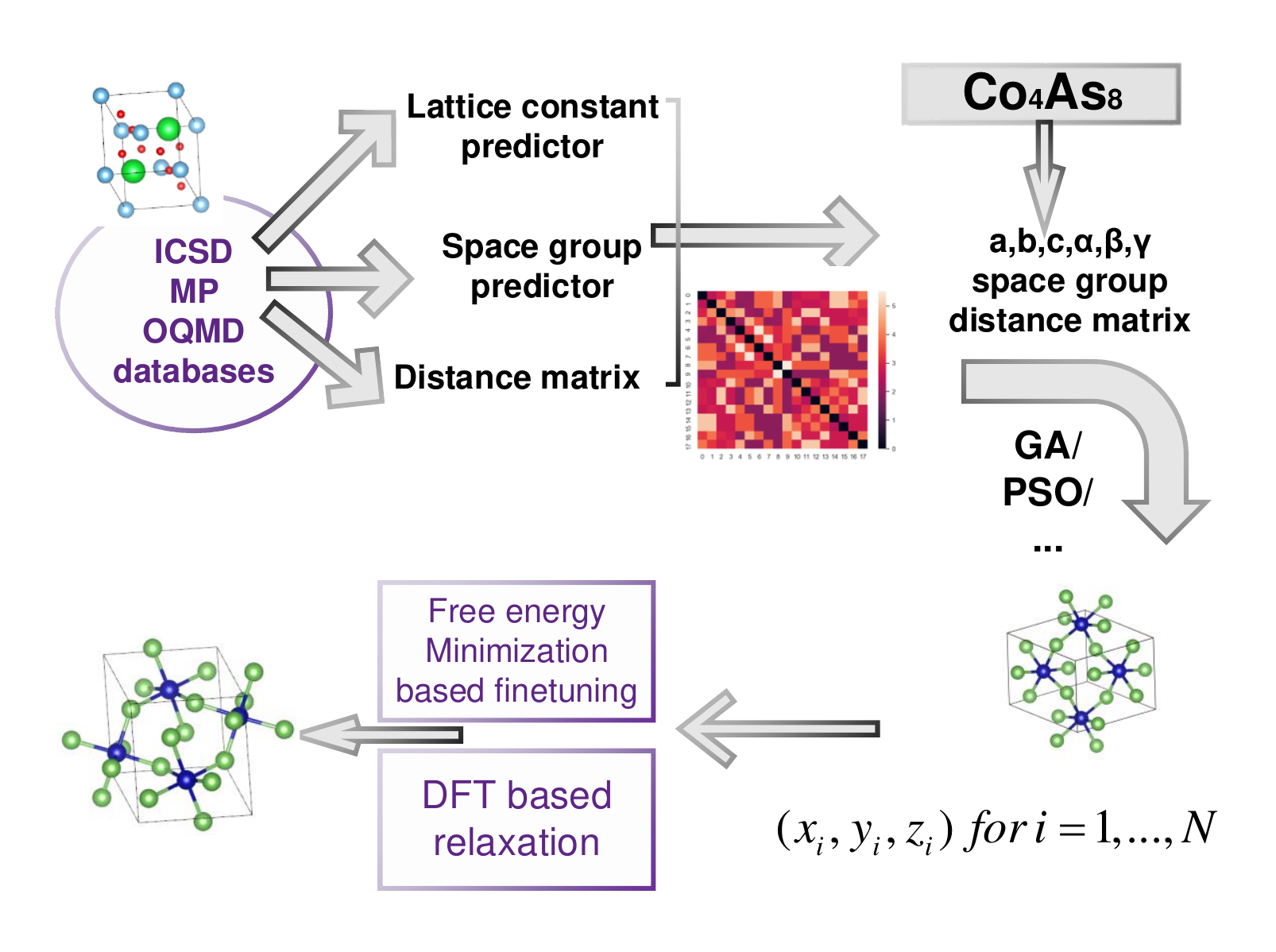}
  \caption{The DMCrystal algorithm for distance matrix based crystal structure prediction. 
 }
  \label{fig:framework}
\end{figure}

\subsection{Distance matrix based crystal structure prediction using global optimization}
In our problem formulation, the independent variables are a set of fractional coordinates $(x_i,y_i,z_i)$ for i=0,...,N, where N is the number of Wyckoff positions and $x_i,y_i,z_i$ are all real numbers in the range of [0,1]. To solve this crystal structure reconstruction problem, we propose to employ global optimization algorithms such as GAs to search the coordinates by maximizing the match between the distance matrix of the predicted structure and the distance matrix of the target crystal structure. This DMCrystal CSP framework is given in Figure~\ref{fig:framework}. Basically, first, using the existing inorganic materials samples in the databases such as ICSD, Materials Project, or OQMD, three prediction models will be trained including a space group predictor \cite{zhao2020machine,liang2020cryspnet}, a lattice constant predictor \cite{zhang2020machine,nait2020prediction,jiang2006prediction,javed2007lattice,majid2010lattice}, and a distance matrix predictor. And then a global optimization algorithm such as the genetic algorithm will be used to search the atom coordinates such that the resulting structure's distance matrix matches the distance matrix of the predicted structure as much as possible. After that, the structures will then be fed to free energy minimization based DFT relaxation or refinement to generate the final structure prediction.


\subsection{Genetic algorithms(GA)}

In this work we focus on exploring how global optimization algorithm such as genetic algorithms can be used to search the atom coordinates guided by a given distance matrix. We evaluate the genetic algorithm on different problem instances. Genetic algorithms \cite{goldberg1988genetic} are population based global optimization algorithms inspired by the biological evolution process. Candidate solutions (individuals) are encoded by binary or real-valued vectors. Starting with a random population of individuals, the population is then subject to generations of mutation, crossover, and selection to evolve the population toward individuals with high fitness, evaluated by the optimization objective functions. Compared to other heuristic search algorithms, GAs have proved to be suitable for large-scale global optimization problems \cite{whitley2019next} and has been used in several crystal structure prediction algorithms \cite{glass2006uspex,curtis2018gator,avery2019xtalopt}, and mainly for free energy minimization. The main hyper-parameters include the population size, crossover and mutation rates, and the number of generations. Here we apply the real-value encoded GA as the global optimization procedure for crystal structure reconstruction from distance matrix. Genetic algorithms also have the flexibility to achieve mixed variable-type optimization.

\subsection{Objective function and Evaluation Criteria}

The objective function for distance matrix based structure reconstruction is defined as the Euclidean distance of two distance matrices, which is shown in the following equation:

\begin{equation}
\operatorname{fitness}_{opt}=\sqrt{\frac{1}{N^2} \sum_{i=1}^{N}\sum_{j=1}^{N}\left\|A_{i,j}-B_{i,j}\right\|^{2}}
\end{equation}

where$A$ is the distance matrix of the candidates of the search algorithm and $B$ is the predicted distance matrix of a given composition, and N is the number of atom in the unit cell.

To evaluate the reconstruction performance of different algorithms, we can use the RMSD error of distance matrices as one evaluation criterion, which however does not indicate the final structure similarity between the predicted structure and the true target structure. To address this, we define the root mean square distance (RMSD) and mean absolute error (MAE) of two structures as below:
\begin{equation}
    \begin{aligned}
\mathrm{RMSD}(\mathbf{v}, \mathbf{w}) &=\sqrt{\frac{1}{n} \sum_{i=1}^{n}\left\|v_{i}-w_{i}\right\|^{2}} \\
&=\sqrt{\frac{1}{n} \sum_{i=1}^{n}\left(\left(v_{i x}-w_{i x}\right)^{2}+\left(v_{i y}-w_{i y}\right)^{2}+\left(v_{i z}-w_{i z}\right)^{2}\right)}
\end{aligned}
\end{equation}

\begin{equation}
        \begin{aligned}
\mathrm{MAE}(\mathbf{v}, \mathbf{w}) &=\frac{1}{n} \sum_{i=1}^{n}\left\|v_{i}-w_{i}\right\| \\
&=\frac{1}{n} \sum_{i=1}^{n}\left(\|v_{i x}-w_{i x}\|+\|v_{i y}-w_{i y}\|+\|v_{i z}-w_{i z}\|\right)
\end{aligned}
\end{equation}

where $n$ is the number of independent atoms in the target crystal structure. For symmetrized cif structures, $n$ is the number of independent atoms of the set of Wyckoff equivalent positions. For regular cif structures, it is the total number of atoms in the compared structure. $v_i$ and $w_i$ are the corresponding atoms in the predicted crystal and the target crystal structure. It should be pointed out that in the experiments of this study, the only constraints for the optimization is the distance matrix, it is possible that the predicted atom coordinates are oriented differently from the target atoms in terms of of coordinate systems. To avoid this complexity, we compare the RMSD and MAE for all possible coordinate systems matching such as (x,y,z -->x,y,z), (x,y,z -->x,z,y), etc. and report the lowest RMSD and MAE. 

We also use the contact map accuracy to evaluate how well a CSP algorithm can reconstruct the topology of the target material. It is defined as the dice coefficient, which is shown in the following equation:
\begin{equation}
\operatorname{contact map accuracy}=\operatorname{Dice}=\frac{2|A \cap B|}{|A|+|B|} \approx\frac{2 \times A \bullet B}{\operatorname{Sum}(A)+\operatorname{Sum}(B)}
\end{equation}

where$A$ is the predicted contact map matrix and $B$ is the true contact map of a given composition, both only contain 1/0 entries.  $A \cap B$  denotes the common elements of A and B, |g| represents the number of elements in a matrix, • denotes dot product, Sum(g) is the sum of all matrix elements. Dice coefficient essentially measures the overlap of two matrix samples, with values ranging from 0 to 1 with 1 indicating perfect overlap. We also call this performance measure as contact map accuracy. 

\section{Experiments}

\subsection{Test problems}

We have selected a set of target crystal structures as test cases for evaluating the proposed distance matrix based crystal structure reconstruction algorithm using a genetic algorithm. The list of target materials are shown in Table \ref{table:target_structures}. Here, the numbers of independent atom sites are 2 and 3 corresponding to 6 and 9 number of optimization variables. The space group numbers range from 4 to 61 corresponding to triclinic,monoclinic,orthorhombic structures (More symmetric structures are reported in Section 3.3.3. 

\begin{table}[H] 
\begin{center}
\caption{Statistics of target crystal structures}
\label{table:target_structures}
 \begin{tabular}{|c c c c c c|} 
  \hline
\textbf{Target} & MP\_id & \textbf{No.of  sites} & \textbf{\#Atom in unit cell} & \textbf{Space Group}& \textbf{\#variables} \\ [0.5ex] 
 \hline\hline
Ag\textsubscript{4}S\textsubscript{2}  & mp-560025  & 3  & 6  & 4 & 9  \\  \hline
Bi\textsubscript{4}Se\textsubscript{4}  & mp-1182022 & 2  & 8  & 14  & 6  \\  \hline
B\textsubscript{4}N\textsubscript{4}  & mp-569655 & 2 & 8   & 14 & 6 \\  \hline
S\textsubscript{4}N\textsubscript{4}  & mp-236 & 2 & 8  & 14 & 6  \\  \hline
Pb\textsubscript{4}O\textsubscript{4}  & mp-550714 & 2  & 8  & 29 & 6  \\  \hline
Co\textsubscript{4}As\textsubscript{8} & mp-2715   & 3  & 12   & 14   & 9   \\  \hline
Bi\textsubscript{8}Se\textsubscript{4} & mp-1102082  & 3  & 12 & 14  & 9   \\  \hline
Te\textsubscript{4}O\textsubscript{8} & mp-561224 & 3 & 12  & 19  & 9  \\  \hline
W\textsubscript{4}N\textsubscript{8}  & mp-754628 & 3  & 12  & 33  & 9  \\  \hline
Cd\textsubscript{4}P\textsubscript{8} & mp-402  & 3   & 12    & 33   & 9   \\  \hline
Ni\textsubscript{8}P\textsubscript{8} & mp-27844 & 2   & 16  & 61  & 6  \\  \hline

\end{tabular}
\end{center}
\end{table}

\subsection{Experimental Setup}



For all optimization algorithms, we set the lower boundary and upper boundary of all variables to be [0, 1] when optimizing fractional coordinates. The number of variables depends on the target materials, which is equal to the number of independent atom sites multiplied by 3. For GA, we set the population size to 100 and the number of generations to 1000 with cross-over probability of 0.8 mutation probability of 0.001.

\subsection{Results}

\begin{figure}[hbt!]
	\centering
	\begin{subfigure}{.4\textwidth}
		\includegraphics[width=\textwidth]{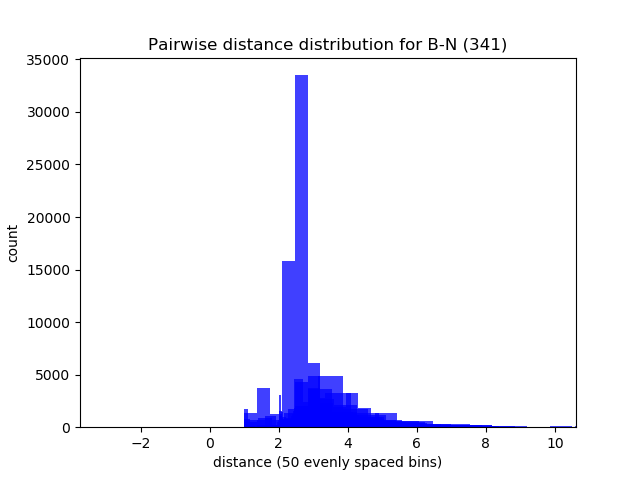}
		\caption{Atom pairs of B and N}
		\vspace{3pt}
	\end{subfigure}
	\begin{subfigure}{.4\textwidth}
		\includegraphics[width=\textwidth]{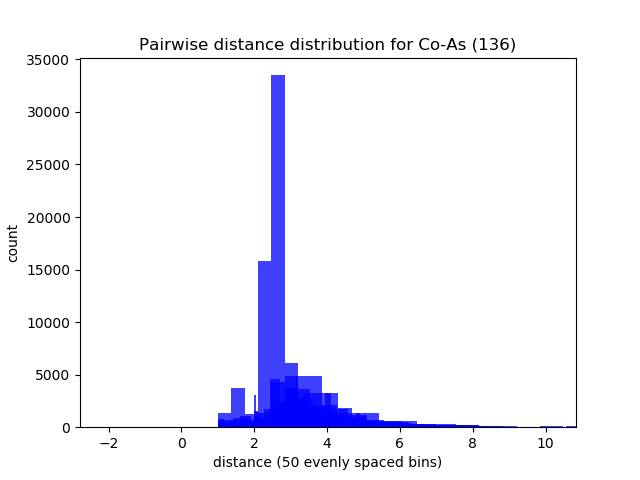}
		\caption{Atom pairs of Co and As }
		\vspace{3pt}
	\end{subfigure}
	\begin{subfigure}{.4\textwidth}
		\includegraphics[width=\textwidth]{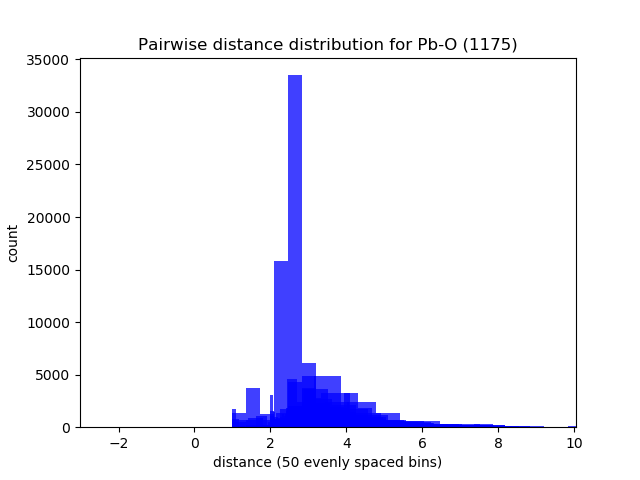}
		\caption{Atom pairs of Pb and O }
	\end{subfigure}
	\begin{subfigure}{.4\textwidth}
		\includegraphics[width=\textwidth]{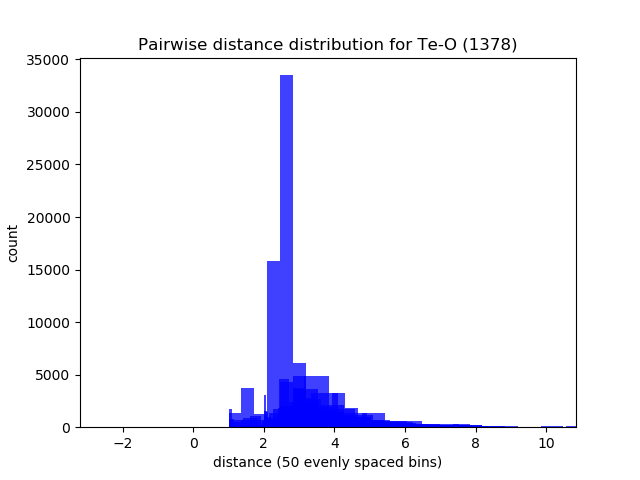}
		\caption{Atom pairs of Te and O }
	\end{subfigure}
	\caption{Pairwise atom distance distribution of atom pairs in ICSD. The highly conserved atom-pair distances mean that many of the atom pair distances may be predicted reliably. }
	\label{distanceHistogram}
\end{figure}

\subsubsection{Distribution analysis of pairwise atomic distances}

One of the main assumptions of our distance matrix based crystal structure prediction is that the distance matrix of a new material with only composition information can be predicted. To show that the pairwise atom distance of a given pair of elements (A,B) can be predicted, we plot the histogram of the shortest distances of these atoms in all known crystals of ICSD structures including B-N and Co-As pairs which both have 341 and 136 pairs respectively. We also include the Pb-O and Te-O atom pairs which have 1175 and 1378 pairs in ICSD structures. From Figure~\ref{distanceHistogram}, we can find that for atom pairs with both high frequencies and low frequencies, the pairwise atom distances are highly conserved, with two very high peaks in all cases. This means that by the two pairs of element information alone, one can achieve good distance prediction, which can be further improved by using more context information from the formula such as other elements for ternary materials.

\subsubsection{Successful distance matrix based crystal structure prediction/reconstruction}

\begin{table}[!htb] 
\begin{center}
\caption{Performances of DMcrystal in terms of structure prediction accuracy}
\label{table:overall_performance}
\begin{tabular} {|l|l|l|l|l|l|l|l|l|l|}
\hline
                    & \multicolumn{3}{c|}{Contact map(CMCrystal)}                                                      & \multicolumn{3}{c|}{Distance matrix   (DMCrystal)}                                               & \multicolumn{3}{c|}{Improvement \%}                                                              \\ \hline
material           & \multicolumn{1}{c|}{\makecell{contact \\map\\ accuracy}} & \multicolumn{1}{c|}{RMSD} & \multicolumn{1}{c|}{MAE} & \multicolumn{1}{c|}{\makecell{contact \\map \\accuracy}} & \multicolumn{1}{c|}{RMSD} & \multicolumn{1}{c|}{MAE} & \multicolumn{1}{c|}{\makecell{contact\\ map\\ accuracy}} & \multicolumn{1}{c|}{RMSD} & \multicolumn{1}{c|}{MAE} \\ \hline
\makecell{Ag\textsubscript{4}S\textsubscript{2}\\(mp-560025)}   & 1.000                                     & 0.320                     & 0.233                    &                 1.000                          & 0.076                     & 0.057                    &                                  0\%         & 76.19\%                   & 75.57\%                  \\ \hline
\makecell{Bi\textsubscript{4}Se\textsubscript{4}\\(mp-1182022)} & 1.000                                     & 0.124                     & 0.097                    &                 1.000                         & 0.069                     & 0.057                    &                                  0\%         & 44.64\%                   & 41.66\%                  \\ \hline
\makecell{B\textsubscript{4}N\textsubscript{4}\\(mp-569655)}    & 1.000                                     & 0.070                     & 0.054                    &                 1.000                          & 0.022                     & 0.017                    &                                  0\%         & 68.03\%                   & 68.94\%                  \\ \hline
\makecell{Pb\textsubscript{4}O\textsubscript{4}\\(mp-550714)}   & 1.000                                     & 0.246                     & 0.196                    &                1.000                           & 0.097                     & 0.078                    &                                 0\%          & 60.48\%                   & 60.30\%                  \\ \hline
\makecell{S\textsubscript{4}N\textsubscript{4}\\(mp-236)}       & 1.000                                     & 0.156                     & 0.137                    &                1.000                           & 0.081                     & 0.064                    &                                 0\%          & 47.93\%                   & 53.16\%                  \\ \hline
\makecell{Te\textsubscript{4}O\textsubscript{8}\\(mp-561224)}   & 1.000                                     & 0.379                     & 0.266                    &                1.000                           & 0.161                     & 0.135                    &                                 0\%          & 57.47\%                   & 49.23\%                  \\ \hline
\makecell{W\textsubscript{4}N\textsubscript{8}\\(mp-754628)}    & 1.000                                     & 0.368                     & 0.214                    &                1.000                           & 0.149                     & 0.130                    &                                 0\%          & 59.41\%                   & 39.11\%                  \\ \hline
\makecell{Cd\textsubscript{4}P\textsubscript{8}\\(mp-402)}      & 1.000                                     & 0.320                     & 0.204                    &                1.000                           & 0.129                     & 0.105                    &                                 0\%          & 59.67\%                   & 48.23\%                  \\ \hline
\makecell{Co\textsubscript{4}As\textsubscript{8}\\(mp-2715)}    & 0.923                                     & 0.197                     & 0.149                    &                1.000                           & 0.229                     & 0.125                    &                                8.3\%           & -16.34\%                  & 16.15\%                  \\ \hline
\makecell{Bi\textsubscript{8}Se\textsubscript{4}\\(mp-1102082)} & 0.889                                     & 0.257                     & 0.232                    &                1.000                           & 0.102                     & 0.083                    &                                12.5\%             & 60.33\%                   & 64.23\%                  \\ \hline
\makecell{Ni\textsubscript{8}P\textsubscript{8}\\(mp-27844)}    & 1.000                                     & 0.381                     & 0.335                    &                0.947                           & 0.128                     & 0.104                    &                               -5.26\%            & 66.53\%                   & 69.01\%                  \\ \hline
\end{tabular}
\end{center}
\end{table}

To evaluate our DMCrystal method for crystal structure prediction, we apply it to a selected set of 11 target structures as shown in Table~\ref{table:target_structures} with the number of atoms ranging from 6 to 16. The total number of objective evaluations is set as 100,000 or 1000 generations for population size of 100 for GA.
The overall performance of of the DMCrystal algorithm for distance matrix based crystal structure reconstruction is shown in Table~\ref{table:overall_performance}, which also includes the contact map based prediction results for comparison. We find that the contact prediction accuracy for 9 out of the 11 targets reach 100\%, demonstrating the effectiveness of our method to find the target topology from random atom coordinates using the distance matrix as the optimization target. This performance scores are similar to those of CMCrystal which directly uses the contact map as the optimization objective \cite{hu2020contact}. Table~\ref{table:overall_performance} also shows the RMSD and MAE scores of the predicted structures compared to the target structures by CMCrystal and DMCrystal, both of which are calculated in terms of fractional coordinates of the independent atom sites. The RMSD values of DMCrystal range from 0.022 for B\textsubscript{4}N\textsubscript{4} to 0.229 for Co\textsubscript{4}As\textsubscript{8} while the MAE scores range from 0.017 for B\textsubscript{4}N\textsubscript{4} to 0.135 for Te\textsubscript{4}O\textsubscript{8}. We found both scores of DMCrystal are much better than those of the CMCrystal algorithm. Indeed, when we compare the RMSD and MAE scores of these two methods for the 11 targets in columns 3/4 and 6/7 for CMCrystal and DMCrystal respectively, we find that for the 11 targets, the DMCrystal method has achieved on average 53\% improvement in terms of RMSD/MAE reconstruction errors compared to CMCrystal algorithm. Out of the 11 target, DMCrystal works slightly worse for only one target structure (Co\textsubscript{4}As\textsubscript{8}) with RMSD of 0.229 versus 0.197 of CMCrystal. But its corresponding MAE of 0.125 is better than the RMSD (0.149) of CMCrystal. In terms of contact map accuracy, the DMCrystal achieves equal or better results for 10 out of 11 targets except Ni\textsubscript{8}P\textsubscript{8}, which is 5.26\% worse. We also observe that from first row Ag\textsubscript{4}S\textsubscript{2} to last row Ni\textsubscript{8}P, the RMSD errors increase with the increasing number of atoms in the structures. These results demonstrate that distance matrix based crystal structure reconstruction/prediction can be much better, this is consistent with the conclusion in protein structure prediction in which the distance matrix based prediction algorithm achieved better performance for contact map based protein structure prediction. The better RMSD/MAE performance scores of DMCrystal is understandable as CMCrystal only aims to reconstruct the topology (the contact map matrix). In Discussion Section, we explore whether adding a few atomic pairwise distance information can improve the performance of CMCrystal. 

\begin{figure}[h!]
	\centering
	\begin{subfigure}{.4\textwidth}
		\includegraphics[width=\textwidth]{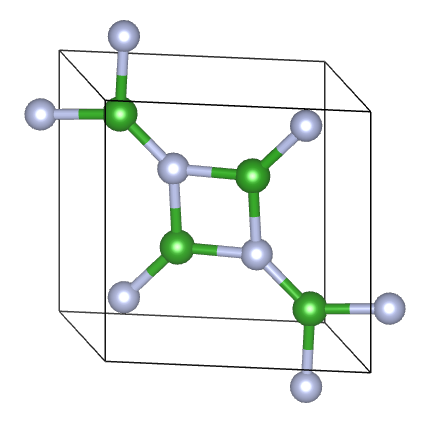}
		\caption{Target structure B\textsubscript{4}N\textsubscript{4}}
		\vspace{3pt}
	\end{subfigure}
	\begin{subfigure}{.4\textwidth}
		\includegraphics[width=\textwidth]{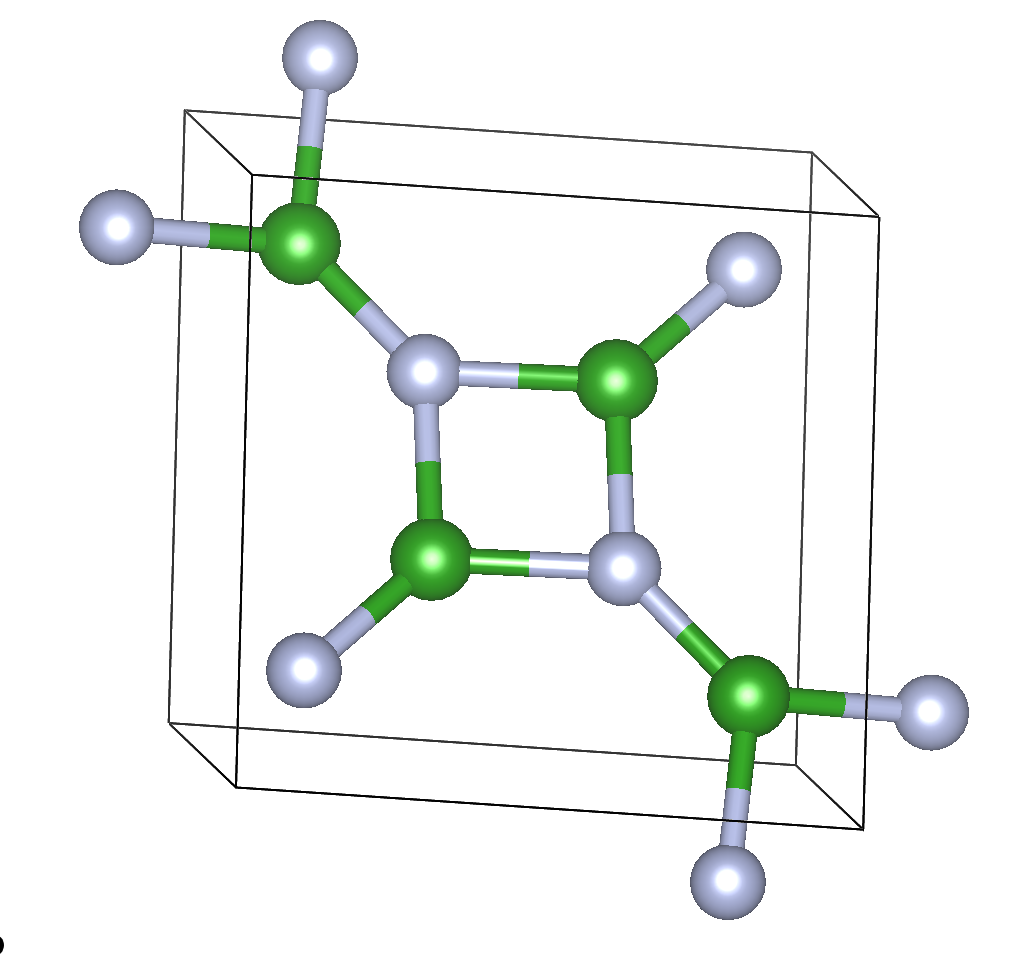}
		\caption{Predicted structure of B\textsubscript{4}N\textsubscript{4} with contact map accuracy:100\%, RMSD:0.022 }
		\vspace{3pt}
	\end{subfigure}
	\begin{subfigure}{.4\textwidth}
		\includegraphics[width=\textwidth]{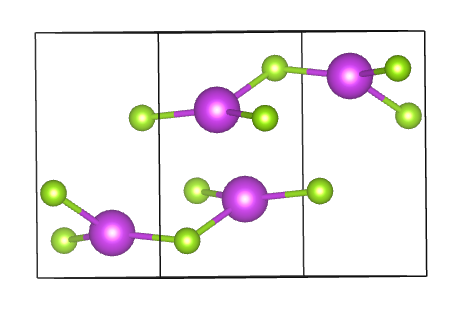}
		\caption{Target structure of Bi\textsubscript{4}Se\textsubscript{4}}
	\end{subfigure}
	\begin{subfigure}{.4\textwidth}
		\includegraphics[width=\textwidth]{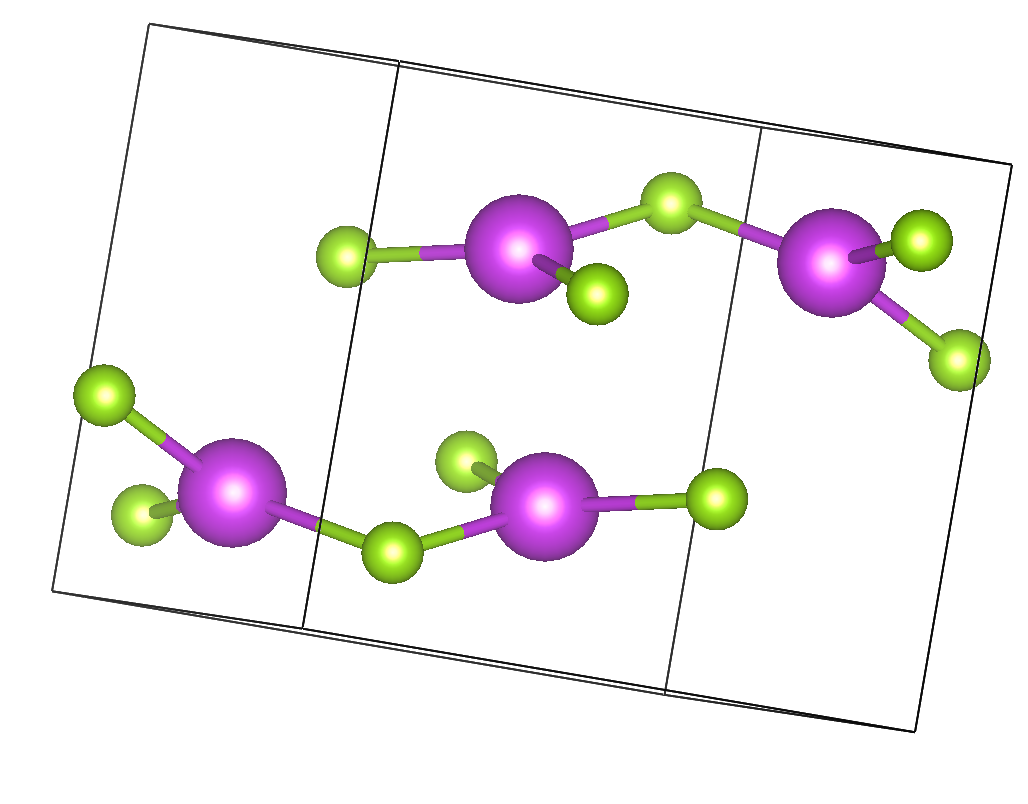}
		\caption{Predicted structure of Bi\textsubscript{4}Se\textsubscript{4} with contact map accuracy:100\%, RMSD:0.069 }
	\end{subfigure}
	
	\begin{subfigure}{.4\textwidth}
		\includegraphics[width=\textwidth]{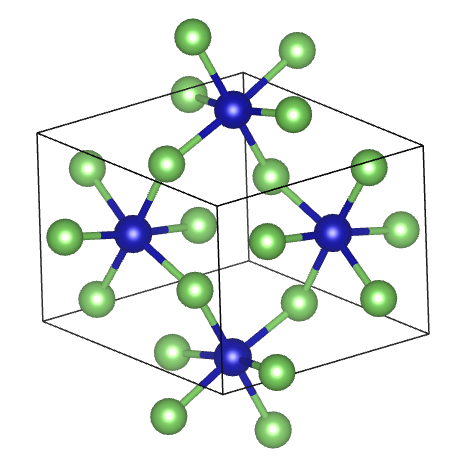}
		\caption{Target structure of Co\textsubscript{4}As\textsubscript{8}}
	\end{subfigure}
	\begin{subfigure}{.4\textwidth}
		\includegraphics[width=\textwidth]{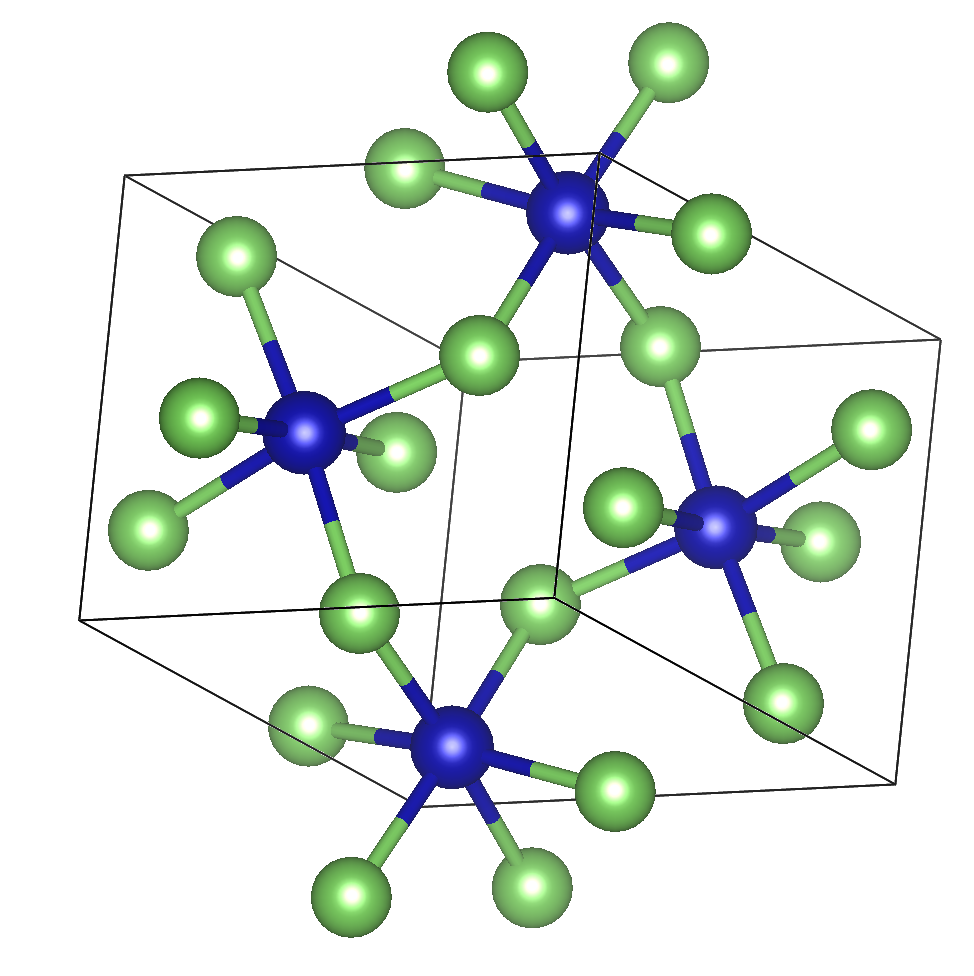}
		\caption{Predicted structure of Co\textsubscript{4}As\textsubscript{8} with contact map accuracy:100\%, RMSD:0.229}
	\end{subfigure}
	
	\caption{Examples of predicted versus target crystal structures by DMCrystal.}
	\label{fig:predictedstructures}
\end{figure}

Figure~\ref{fig:predictedstructures}  shows three sets of predicted and target crystal structures of B\textsubscript{4}N\textsubscript{4}, Bi\textsubscript{4}Se\textsubscript{4}, and Co\textsubscript{4}As\textsubscript{8}. For both B\textsubscript{4}N\textsubscript{4} and Bi\textsubscript{4}Se\textsubscript{4} (Figure~\ref{fig:predictedstructures} (a)-(d)), the contact map accuracy reaches 100\% and the predicted structures are very close to the target structures. The RMSD of B\textsubscript{4}N\textsubscript{4} is 0.022 which is smaller than the RMSD (0.124) of Bi\textsubscript{4}Se\textsubscript{4}, which is reflected by the higher similarity of the pairs of B\textsubscript{4}N\textsubscript{4} than the pair of structures of Bi\textsubscript{4}Se\textsubscript{4}. The prediction accuracy for target structure of Co\textsubscript{4}As\textsubscript{8} is lower with RMSD of 0.229. Compared to CMCrystal, we find that the DMCrystal algorithm usually can reach much more similar structures to the true structures due to the explicit distance constraints while both algorithms can achieve high accuracy in topology prediction.

\FloatBarrier

\subsubsection{Performance comparison of different reconstruction algorithms using distance information}

\begin{figure}[ht!]
  \centering
  \includegraphics[width=0.8\linewidth]{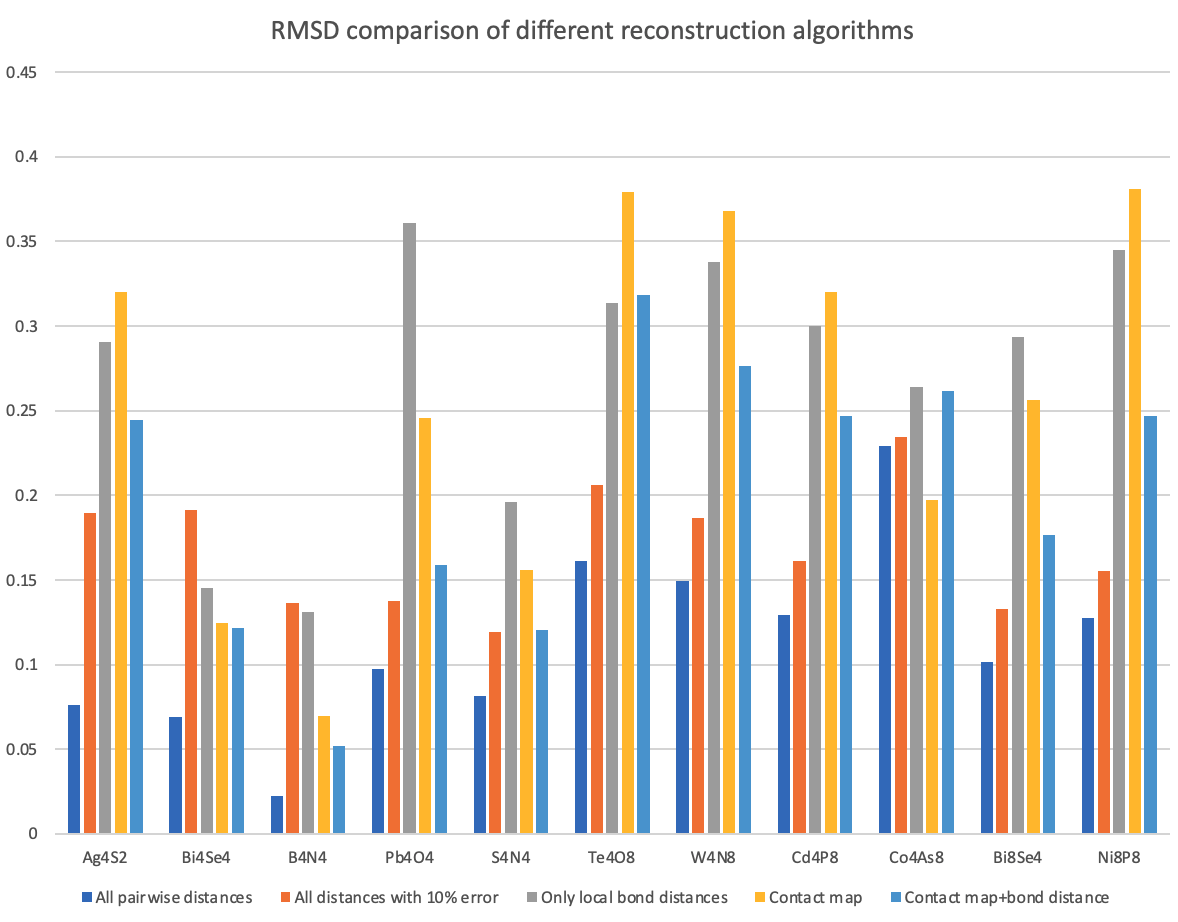}
  \caption{Performance comparison of different algorithms in terms of structure reconstruction RMSD. }
  \label{fig:algorithmcomparison}
\end{figure}

The crystal structure prediction performance of our DMCrystal algorithms depends on the quality of the predicted distance matrix. In a realistic crystal structure prediction, the predicted distance matrices may have errors, especially for those non-local non-bond distances. Here we investigate how the different types of distance information may be used for crystal structure prediction using either DMCrystal or CMCrystal frameworks. For comparison purpose, we evaluate the following five algorithms sorted by decreasing amount of distance information used: 1) DMCrystal, which use all atom pair distance information; 2) DMCrystal with 10\% noise on the distance matrix; 3) DMCrystal with only local bond distance information; 4) CMCrystal with local bond distance information; 5) CMCrystal: only contact map is used with no distance information. We ran all the algorithms over the 11 target structures in Table~\ref{table:target_structures}. The RMSD performance scores of all five algorithms are shown in Figure~\ref{fig:algorithmcomparison}

First, we find that for all 11 targets, the DMCrystal achieves the lowest RMSD errors due to its exploitation of all pairwise atom distance information. When we add 10\% of noise to the distance matrix, the performance degrades accordingly, but still better than the performance of the reconstruction algorithms using only local bond distances. The contact map based reconstruction algorithm achieves the worst performance in terms of RMSD for five out of 11 targets, reflecting its focus on topology(contact map) rather than the RMSD. However, its performance can be improved by adding the local bond distance information, which are usually easy to be predicted due to their high conservation (see Figure~\ref{distanceHistogram}). 

\subsubsection{Comparison with existing crystal structure prediction algorithms}

To show the potential advantages of our knowledge-rich distance matrix based crystal structure prediction method compared to existing free energy minimization based ab initio method, we also apply CALYPSO \cite{CALYPSO} to the three targets shown in Figure~\ref{fig:predictedstructures}. And then we calculate the formation energies of all their predicted structures as well of three predicted structures by the DMCrystal algorithm using density functional theory (DFT). DFT calculations are carried out based on the Vienna ab initio simulation package (VASP) \cite{Vasp4,hafner2008ab}. The interactions between the ions and electrons are treated by the projected augmented wave (PAW) method \cite{blochl1994projector,kresse1999ultrasoft}. The generalized gradient approximation (GGA) based exchange-correlation potential is considered with the Perdew-Burke-Ernzerhof (PBE) pseudopotentials \cite{perdew1996generalized_GGA1}. A plane-wave energy cutoff is set to 500 eV.  The convergence criteria for total energy and the Hellmann-Feyman forces are $1.0\times 10^{-6}$ eV and 0.01eV/ Å, respectively. The $\Gamma$-centered Monkhorst-Pack k-meshes are considered. The formation energies of the materials are computed based on the following equation. 
\begin{equation}
    E_{form} = \frac{1}{(x+y+z)}(E_{\textrm{A}_x \textrm{B}_y \textrm{C}_z }-x\mu_\textrm{A}- y\mu_\textrm{B}-z\mu_\textrm{C}),
\end{equation}
 where $E_{\textrm{A}_x \textrm{B}_y \textrm{C}_z }$ is the total energy of the material with the chemical formula $\textrm{A}_x \textrm{B}_y \textrm{C}_z $ and  $\mu$ represents the energy of each atom in its respective ground-state bulk phase \cite{Siriwardane2020}.

For CALYPSO, we let it run 24 hours on a 8-core Dell server with 3.7GHz multi-core CPU. However, we are not able to achieve successful runs when specifying the space group for CALYPSO runs. So these experiments for CALYPSO have not specified the target space group, thus leading to predicted structures that may match the targets in terms of atom site numbers. We also run DMCrystal on the same machine but using only one core, which finishes each of the three running experiments for the three targets B\textsubscript{4}N\textsubscript{4}, Bi\textsubscript{4}Se\textsubscript{4}, Co\textsubscript{4}As\textsubscript{8} within 45 minutes.

\begin{figure}[h!]
	\centering
	\begin{subfigure}{.37\textwidth}
		\includegraphics[width=\textwidth]{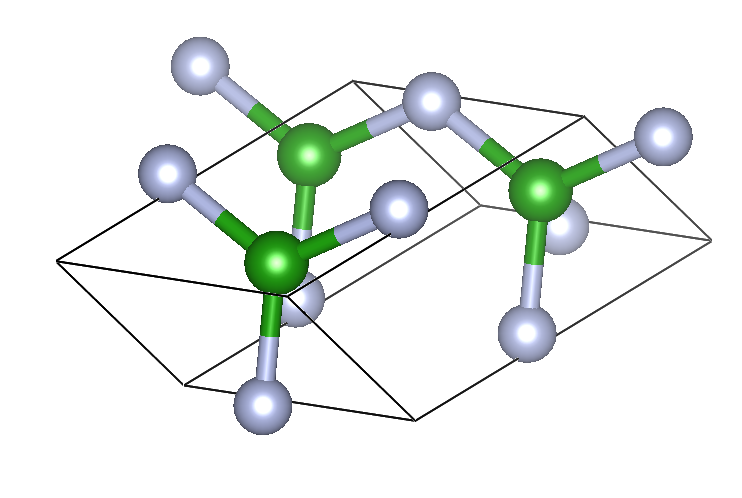}
		\caption{Predicted structure of B\textsubscript{4}N\textsubscript{4} with RMSD:N/A. Final atom sites different from target structure}
		\vspace{3pt}
	\end{subfigure}
	\begin{subfigure}{.3\textwidth}
		\includegraphics[width=\textwidth]{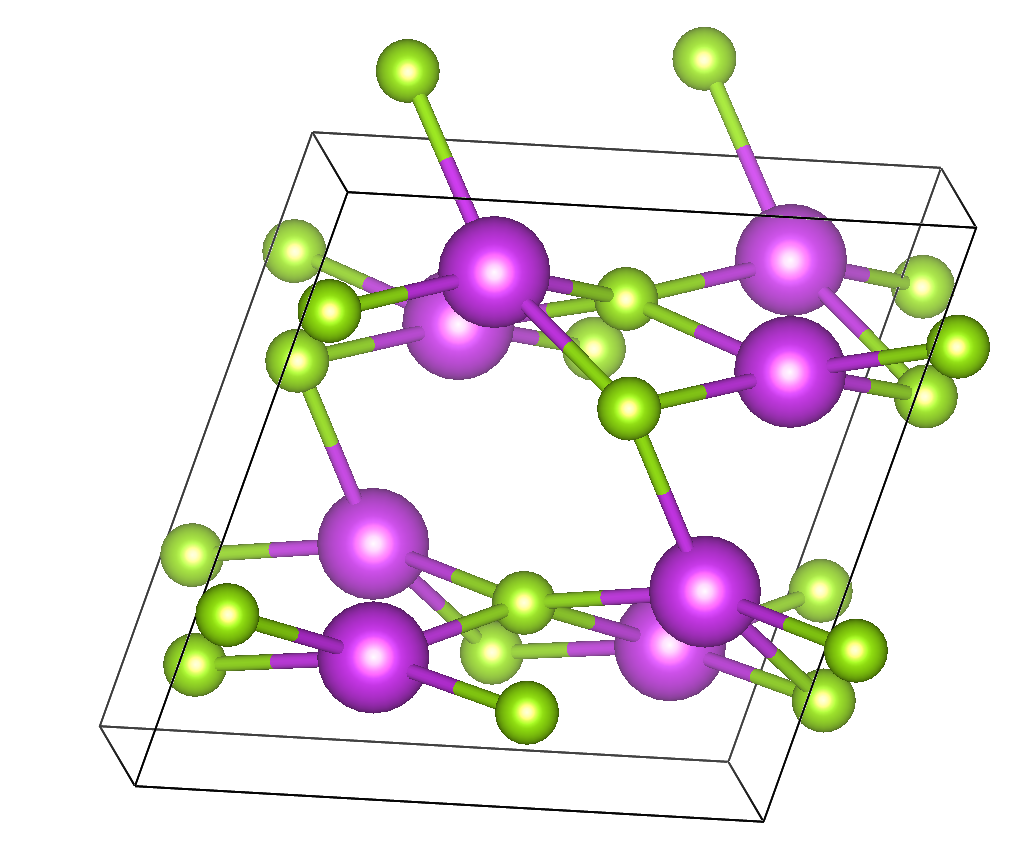}
		\caption{Predicted structure of Bi\textsubscript{4}Se\textsubscript{4} with RMSD:N/A. Final atom sites different from target structure  }
		\vspace{3pt}
	\end{subfigure}
	\begin{subfigure}{.3\textwidth}
		\includegraphics[width=\textwidth]{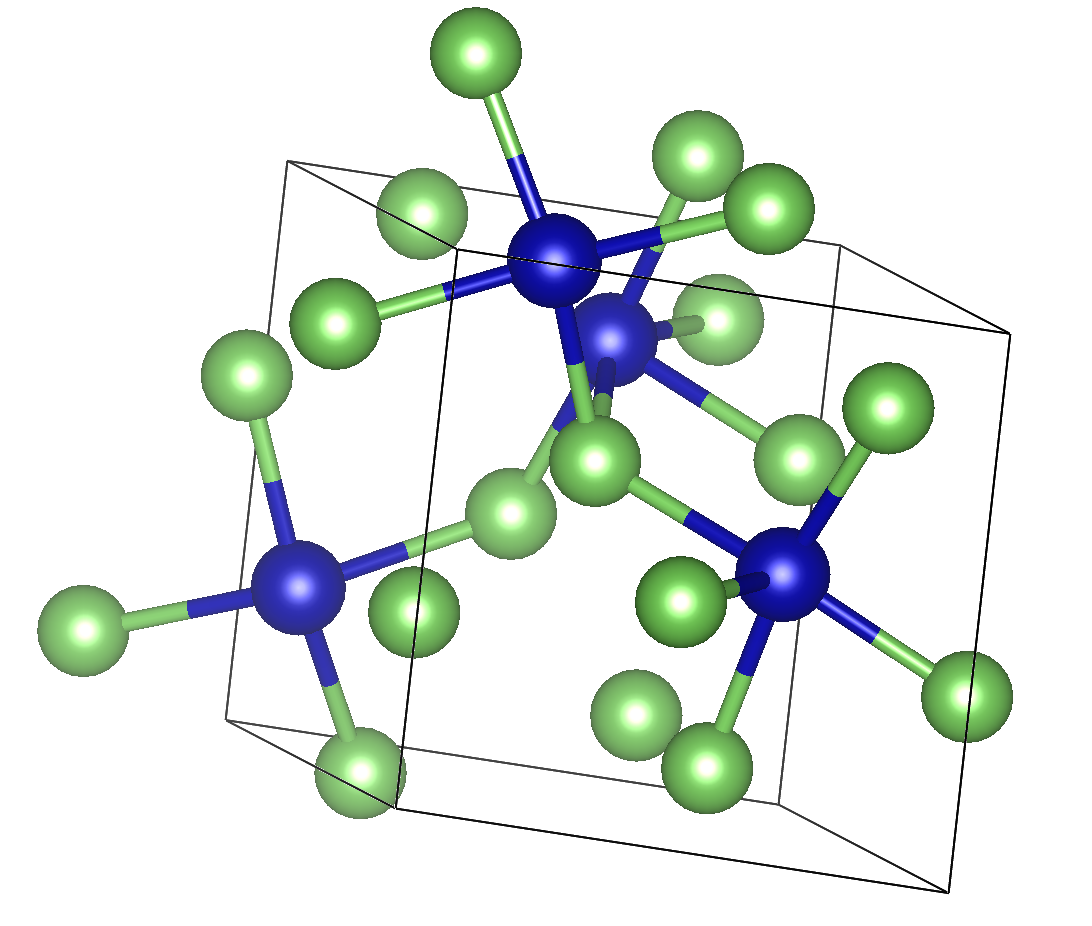}
		\caption{Predicted structure of Co\textsubscript{4}As\textsubscript{8} with RMSD: 0.319}
	\end{subfigure}
	\caption{Structures of three targets predicted by CALYPSO after 24 hours running on a 8-core Dell server.}
	\label{fig:calypsoresults}
\end{figure}

First, Figure~\ref{fig:calypsoresults} shows the predicted structures by CALYPSO for the three benchmark target structures. For the first two targets, CALYPSO found structures with different number of sites so that no RMSE errors can be calculated. For Co\textsubscript{4}As\textsubscript{8}, the RMSD is 0.319 compared to 0.229 by DMCrystal. When comparing the predicted structures by CALYPSO to those predicted by DMCrystal in Figure~\ref{fig:predictedstructures}, we can find that the latter are much similar to the true target structures, which is reflected by the much lower RMSD scores. We have also compared the formation energy of the predicted structures by CALYPSO and DMCrystal as shown in Table~\ref{table:performancecompare}. All three structures predicted by DMCrystal has negative formation energy values (Row 3). With DFT relaxations, all three predicted structures get lower free energy values with Bi\textsubscript{4}Se\textsubscript{4} almost having 50\% reduction in free energy from -0.110 to -0.206 eV/atom. The fine tuned structure has similar free energy with the result by expensive CALYPSO result which however, does not match the target structure. For B\textsubscript{4}N\textsubscript{4}, CALYPSO result has lower formation energy, but the structure does not match the target. For Co\textsubscript{4}As\textsubscript{8}, DMCrystal result has lower formation energy of -0.320 ev/atom compared to the -0.161 ev/atom of the structure predicted by CALYPSO. Since our CALYPSO runs have no space group information, the comparison is not a stringent performance comparison, but it shows the potential of our DMCrystal for crystal structure reconstruction from distance matrices in terms of both speed and quality. 

\begin{table}[!htb] 
\scriptsize
\begin{center}
\caption{ Prediction performance comparison with CALYPSO }
\label{table:performancecompare}
\begin{tabular}{|l|l|l|l|l|}
\hline
 & & \multicolumn{3}{l|}{Formation Energy (eV/atom)} \\ \hline
Method & Running time & B\textsubscript{4}N\textsubscript{4} & Bi\textsubscript{4}Se\textsubscript{4} & Co\textsubscript{4}As\textsubscript{8} \\ \hline
DMCrystal & 45 min (using 1-core) & -0.909 & -0.110 & -0.319 \\ \hline
DMCrystal + DFT &  & -0.977 & -0.206 & -0.320 \\ \hline
CALYPSO & 24 hours (using 8-core) & -1.2934 & -0.208 & -0.161 \\ \hline
  & & RMSD & RMSD & RMSD\\ \hline
DMCrystal  & & 0.022 & 0.069 & 0.229\\ \hline
 CALYPSO &  & N/A & N/A & 0.319\\ \hline
\end{tabular}
\end{center}
\end{table}

\section{Discussion}

In this work, our main focus is on crystal structure reconstruction from a predicted distance matrix along with the predicted space group and lattice parameters. In our experiments, we have used the real distance matrix, space group, and lattice parameters from the benchmark target structures. However, in reality, all these information are predicted for a given composition or materials formula. 

Machine learning predictions of space groups and crystal systems from composition have been proposed in \cite{liang2020cryspnet, zhao2020machine}. It is reported that  \cite{liang2020cryspnet} that the top three space group prediction performance can reach to a range of 0.81 to 0.98 in terms of $R^2$ scores based on different crystal systems. For cubic structures, the $R^2$ score is 0.96 on average. When evaluating their machine learning performance for 18 space group prediction, the F1 scores can reach 0.787 \cite{zhao2020machine}. These two works show that we can predict the space group reasonably well, at least for top K performance. It is thus possible to predict e.g. top 10 candidate space groups for DMcrystal based structure reconstruction and then select the solutions with lowest formation energy. Composition based crystal lattice parameter prediction methods have also been proposed  \cite{takahashi2017descriptors}, which trained a support vector regressor using 1541 binary body centered cubic crystals and has achieved a $R^2$ score of 83.6\% accuracy via cross-validation and maximum error of 4\%. We have improved this result to  $R^2$ of 0.97 for cubic structures.

We also examined additional factors that affect the structure reconstruction quality based on the target structure properties.
From our extensive experiments, we find that there are several factors that affect the crystal structure prediction quality by our algorithm such as the number of independent atom sites, the number of atoms in the unit cell, and the number of bonds/topology constraints and etc. 
Here we report how two factors, the number of independent atomic sites in the unit cell of the material and its space group, affect the crystal structure reconstruction performance by DMCrystal. In Table~\ref{table:accuracy_atomsiteno} we compare the performance of DMCrystal for target structures with the same number of total atoms in the unit cells but different numbers of independent atom sites. It shows that in general, the contact map accuracy decreases when there are more atom sites in the structure. However the RMSD prediction error are relatively stable around 0.1 for all the structures.

\begin{table}[!htb] 
\begin{center}
\caption{ Prediction performance versus number of atom sites for DMCrystal }
\label{table:accuracy_atomsiteno}
\begin{tabular}{|l|l|l|l|l|}
\hline
\multicolumn{1}{|c|}{Target} & \multicolumn{1}{c|}{mp\_id} & \multicolumn{1}{c|}{atom site} & \multicolumn{1}{c|}{contact map accuracy} & \multicolumn{1}{c|}{RMSD} \\ \hline
La\textsubscript{12}Se\textsubscript{16}                     & mp-491                      & 2                              & 1.000                                     & 0.102                     \\ \hline
Bi\textsubscript{8}Pd\textsubscript{4}O\textsubscript{16}                    & mp-29259                    & 3                              & 1.000                                     & 0.124                     \\ \hline
V\textsubscript{8}O\textsubscript{20}                        & mp-25280                    & 4                              & 1.000                                     & 0.083                     \\ \hline
Fe\textsubscript{12}O\textsubscript{16}                      & mp-1192788                  & 5                              & 0.970                                     & 0.120                     \\ \hline
Ge\textsubscript{8}S\textsubscript{12}I\textsubscript{8}                     & mp-27928                    & 6                              & 0.923                                     & 0.091                     \\ \hline
Li\textsubscript{4}V\textsubscript{4}Si\textsubscript{4}O\textsubscript{16}                  & mp-1176508                  & 7                              & 0.889                                     & 0.149                     \\ \hline
Ba\textsubscript{2}V\textsubscript{8}O\textsubscript{18}                     & mp-18910                    & 8                              & 0.881                                     & 0.091                     \\ \hline
Si\textsubscript{2}H\textsubscript{18}C\textsubscript{6}Cl\textsubscript{2}                  & mp-867818                   & 9                              & 0.857                                     & 0.116                     \\ \hline
Zr\textsubscript{8}Cr\textsubscript{8}F\textsubscript{12}                    & mp-690241                   & 10                             & 0.842                                     & 0.164                     \\ \hline
\end{tabular}
\end{center}
\end{table}

Table~\ref{table:accuracy_spacegroup} shows the prediction results of DMCrystal for a set of materials with five atom sites and similar numbers of atoms but different space groups. First, it can be observed that the higher the space group number, the higher the contact map accuracy, indicating that higher symmetry (with higher space group number) puts more constraints on the atom configurations and reduces the search space so that better performance can be achieved by the genetic algorithm in atom coordinate search. In the table, the contact map accuracy increases from 0.828 to 0.963 when the space group goes from 2 to 194. At the same time, the RMSD errors are relative stable around 0.12 with slight increase for targets of space groups 129 and 194. We also compare these performances with those of CMCrystal which only uses contact map as objectives for crystal structure reconstruction and we find the RMSD errors of DMCrystal is much lower.

\begin{table}[htb!] 
\begin{center}
\caption{ Prediction performance versus space group using DMCrystal}
\label{table:accuracy_spacegroup}
\begin{tabular}{|l|l|l|l|l|l|}
\hline
\multicolumn{1}{|c|}{Target} & \multicolumn{1}{c|}{mp\_id} & \multicolumn{1}{c|}{atom site\#} & \multicolumn{1}{c|}{space group} & \multicolumn{1}{c|}{contact map accuracy} & \multicolumn{1}{c|}{RMSD} \\ \hline
Hg\textsubscript{8}Cl\textsubscript{4}O\textsubscript{4}                     & mp-636805                   & 5                                & 2                                & 0.828                                     & 0.115                     \\ \hline
Be\textsubscript{4}B\textsubscript{2}O\textsubscript{10}                     & mp-1079124                  & 5                                & 5                                & 0.833                                     & 0.149                     \\ \hline
Bi\textsubscript{6}O\textsubscript{8}F\textsubscript{2}                      & mp-757162                   & 5                                & 13                               & 0.875                                     & 0.086                     \\ \hline
Tl\textsubscript{6}V\textsubscript{2}O\textsubscript{8}                      & mp-29047                    & 5                                & 44                               & 0.900                                     & 0.120                     \\ \hline
La\textsubscript{4}Sn\textsubscript{2}S\textsubscript{10}                    & mp-12170                    & 5                                & 55                               & 0.900                                     & 0.132                     \\ \hline
Li\textsubscript{2}V\textsubscript{2}F\textsubscript{12}                     & mp-753573                   & 5                                & 102                              & 0.923                                     & 0.107                     \\ \hline
Yb\textsubscript{4}H\textsubscript{4}O\textsubscript{8}                      & mp-625103                   & 5                                & 113                              & 0.941                                     & 0.083                     \\ \hline
Mg\textsubscript{4}Co\textsubscript{2}H\textsubscript{10}                    & mp-642660                   & 5                                & 129                              & 0.966                                     & 0.158                     \\ \hline
K\textsubscript{10}Cu\textsubscript{2}As\textsubscript{4}                    & mp-14623                    & 5                                & 194                              & 0.963                                     & 0.132                     \\ \hline
\end{tabular}
\end{center}
\end{table}

\FloatBarrier
\section{Conclusion}

We formulate the crystal structure prediction/reconstruction problem based on its space group symmetry and the atom pairwise distance matrix, and proposed a genetic algorithm based global optimization to solve this problem. Our experiments show that the genetic algorithm is able to reconstruct the crystal structure for a set of benchmark materials within 45 minutes by setting the pair-wise atomic distance matrix as the optimization objective given only their space group. lattice parameters and the stoichiometry. These predicted structures are close to the target crystal structures so that they can be used to seed the costly free energy minimization based ab initio crystal structure prediction algorithms for further structure refining. They may also be used for DFT based structure relaxation to obtain the correct crystal structures for some compositions. We also find that distance matrix can lead to much higher quality reconstruction compared to the contact map based crystal structure prediction. Our results show that our distance matrix based approach is a viable crystal structure prediction/reconstruction method, similar to their role in protein structure prediction.


\section{Availability of data}

The data that support the findings of this study are openly available in Materials Project database at http:\\www.materialsproject.org 

\section{Contribution}
Conceptualization, J.H.; methodology, J.H. and W.Y.; software, W.Y. and J.H; validation, W.Y, J.H., D.S.;  investigation, J.H., W.Y.; resources, J.H.; data curation, J.H. and W.Y.; writing--original draft preparation, J.H.; writing--review and editing, J.H, W.Y., D.S.; visualization, J.H. and W.Y; supervision, J.H.;  funding acquisition, J.H.

\section{Acknowledgement}
Research reported in this work was supported in part by NSF under grant and 1940099 and 1905775 and by NSF SC EPSCoR Program under award number (NSF Award OIA-1655740 and GEAR-CRP 19-GC02). The views, perspective, and content do not necessarily represent the official views of the SC EPSCoR Program nor those of the NSF.

\bibliography{references}
\bibliographystyle{unsrt}

\end{document}